\documentclass[12pt,preprint]{aastex}
\usepackage{psfig,epsfig,amsfonts,amsmath,amssymb,graphicx,morefloats,appendix,tensor}
\usepackage{color}
\usepackage{natbib}
\begin{document}

\slugcomment{Submitted to ApJL: 2017 December 22 --  Accepted to ApJL: 2018 February 5}

\title{Detection of a Millimeter Flare From Proxima Centauri
}

\author{Meredith A. MacGregor\altaffilmark{1,2}, Alycia J. Weinberger\altaffilmark{1}, David J. Wilner\altaffilmark{3}, Adam F. Kowalski\altaffilmark{4,5}, Steven R. Cranmer\altaffilmark{4}}

\altaffiltext{1}{Department of Terrestrial Magnetism, Carnegie Institution for Science, 5241 Broad Branch Road NW, Washington, DC 20015, USA}
\altaffiltext{2}{NSF Astronomy and Astrophysics Postdoctoral Fellow}
\altaffiltext{3}{Harvard-Smithsonian Center for Astrophysics, 60 Garden Street, Cambridge, MA 02138, USA}
\altaffiltext{4}{Department of Astrophysical and Planetary Sciences, University of Colorado Boulder, 2000 Colorado Ave, Boulder, CO 80305, USA}
\altaffiltext{5}{National Solar Observatory, University of Colorado Boulder, 3665 Discovery Drive, Boulder, CO 80303, USA}

\begin{abstract}

We present new analyses of ALMA 12-m and ACA observations at 233~GHz (1.3~mm) of the Proxima Centauri system with sensitivities of 9.5 and 47~$\mu$Jy~beam$^{-1}$, respectively, taken from 2017 January 21 through 2017 April 25.  These analyses reveal that the star underwent a significant flaring event during one of the ACA observations on 2017 March 24.  The complete event lasted for approximately 1~minute and reached a peak flux density of $100\pm4$~mJy, nearly a factor of $1000\times$ brighter than the star's quiescent emission.  At the flare peak, the continuum emission is characterized by a steeply falling spectral index with frequency, $F_\nu \propto \nu^\alpha$ with $\alpha = -1.77\pm0.45$, and a lower limit on the fractional linear polarization of $|Q/I| = 0.19\pm0.02$.  Since the ACA observations do not show any quiescent excess emission, we conclude that there is no need to invoke the presence of a dust belt at $1-4$~AU.  We also posit that the slight excess flux density of $101\pm9$~$\mu$Jy observed in the 12-m observations compared to the photospheric flux density
of $74\pm4$ $\mu$Jy extrapolated from infrared wavelengths may be due to coronal heating from continual smaller flares, as is seen for AU Mic, another nearby, well-studied, M dwarf flare star.  If this is true, then the need for warm dust at $\sim0.4$~AU is also removed.

\end{abstract}


\section{Introduction}
\label{sec:intro}

Low mass (i.e., M-type stars) are the most common stars in the Galaxy \cite[e.g.,][]{Henry:2006} and have a high frequency of Earth-sized planets both overall and in the habitable zone \citep{Dressing:2015}. However, there is debate as to what extent planets around M dwarfs would be amenable to developing and sustaining life. Late M-type stars, in particular, have long pre-main sequence phases during which the stellar luminosity can change significantly, as well as high stellar activity throughout their entire lifetimes \cite[e.g.,][]{Hawley:1997,Shkolnik:2014,France:2016}.  

Proxima Centauri has garnered tremendous recent interest due to the radial velocity detection of a potentially Earth-mass planet within the habitable zone \citep{ang16} and a new candidate transit event \citep{Li:2017}. At a distance of only 1.3~pc \citep{van07}, Proxima Centauri b ($m_p\text{sin}i = 1.3 M_\oplus$, $a=0.05$~AU) is the closest extrasolar planet to the Solar System.  However, Proxima Centauri (spectral type M5.5V) has long been known as a flare star \citep[e.g.,][]{Thackeray:1950}. Already, many studies have examined the effect that its variability may have on the properties of its planet \citep{Garraffo:2016,Dong:2017,Lingam:2017,Ribas:2017,Zahnle:2017}.

Here, we present a new analysis of 233~GHz (1.3~mm) observations using the ALMA 12-m array and Atacama Compact Array (ACA) that were originally published in \cite{ang17}.  In \S\ref{sec:obs} we briefly summarize the relevant observations, followed by the results of our analysis for the ALMA 12-m and ACA observations in \S\ref{sec:12m} and \S\ref{sec:aca}, respectively.  We discuss the implications of these results for stellar emission mechanisms in \S\ref{sec:star} and the dust belts proposed by \cite{ang17} in \S\ref{sec:dust}.  In \S\ref{sec:conclusions}, we present our conclusions.

\section{Observations}
\label{sec:obs}

Proxima Centauri was observed with ALMA at 233~GHz (1.3 mm) from 2017 January 21 through 2017 April 25 (PI: Anglada, 2016.A.00013.S), and the observations were published in \cite{ang17}.  Between January and April, the target was observed a total of 15 times, $2\times\sim$1.3~hour scheduling blocks (SBs) using the full 12-m array (41 antennas) and $13\times\sim$1.6~hour SBs using the ACA (8--11 7-m antennas).  During all of these observations, the target was observed in 6.58~minute integrations alternating with observations of a phase calibrator, either J1424--6807 or J1329--5608 ($5\fdg5$ and $10\fdg0$ away from the target, respectively).  Several sources were used for both absolute flux calibration (Ganymede, Callisto, Titan, J1427−4206, and J1517−2422) and bandpass calibration (J1427−4206 and J1924−2914).  Overall, the weather was very good during these observations with precipitable water vapor (PWV) ranging between $0.58-1.74$~mm.  The correlator was set-up to maximize continuum sensitivity with four spectral windows centered at 225, 227, 239 and 241~GHz.  Each  spectral window has an effective bandwidth of 1.875~GHz split into 120 channels for a total bandwidth of $\sim$7.5~GHz.  Additional discussion of these observations can be found in \cite{ang17}.  We obtained the raw data from the ALMA archive and generated calibrated measurement sets in \texttt{CASA} \cite[][version 4.7.0 for the ACA data, and 4.7.2 for the 12-m array]{Petry:2012} using the scripts provided.  All continuum images were produced using the \texttt{clean} task in \texttt{CASA}, which inverts the millimeter visibilities to produce deconvolved images.

\section{Results and Analysis}
\label{sec:results}

\subsection{ALMA 12-m Observations}
\label{sec:12m}

\begin{figure}[htbp]
\begin{minipage}[h]{0.55\textwidth}
  \begin{center}
  \vspace{0.5cm}
       \includegraphics[scale=0.55]{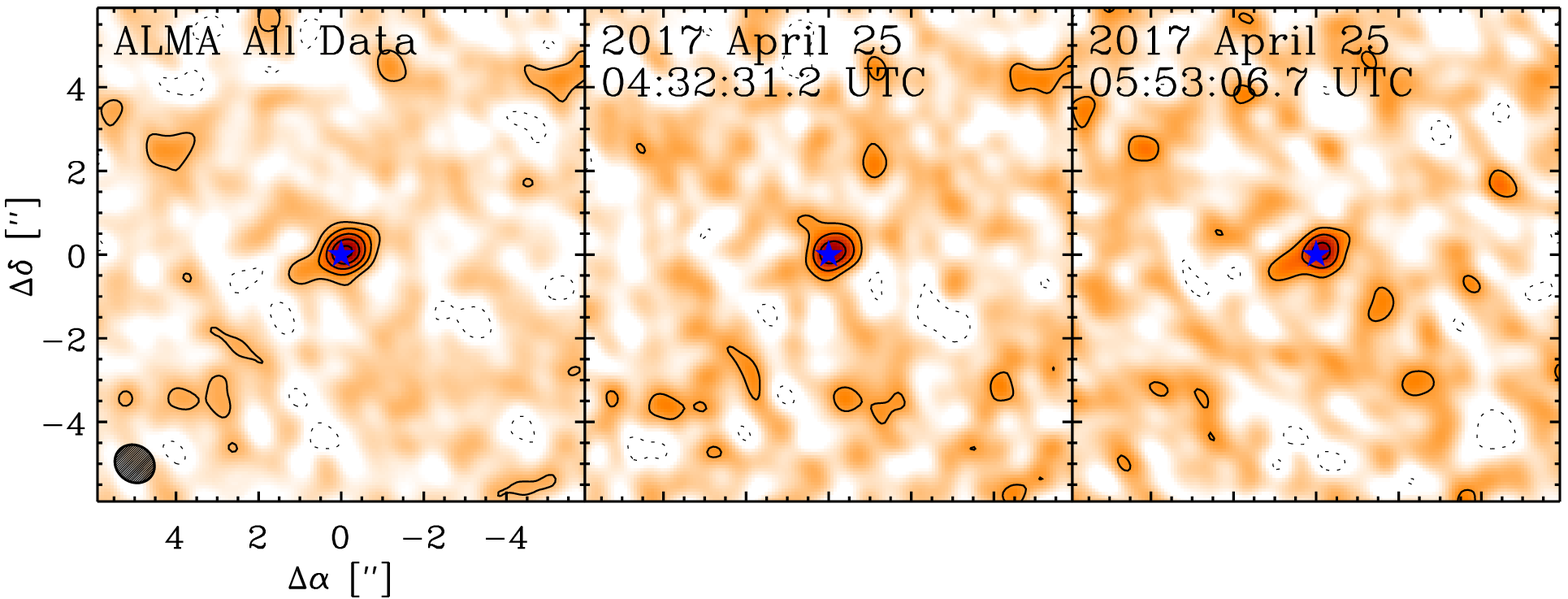}
  \end{center}
 \end{minipage}
\begin{minipage}[h]{0.5\textwidth}
  \begin{center}
       \includegraphics[scale=0.35]{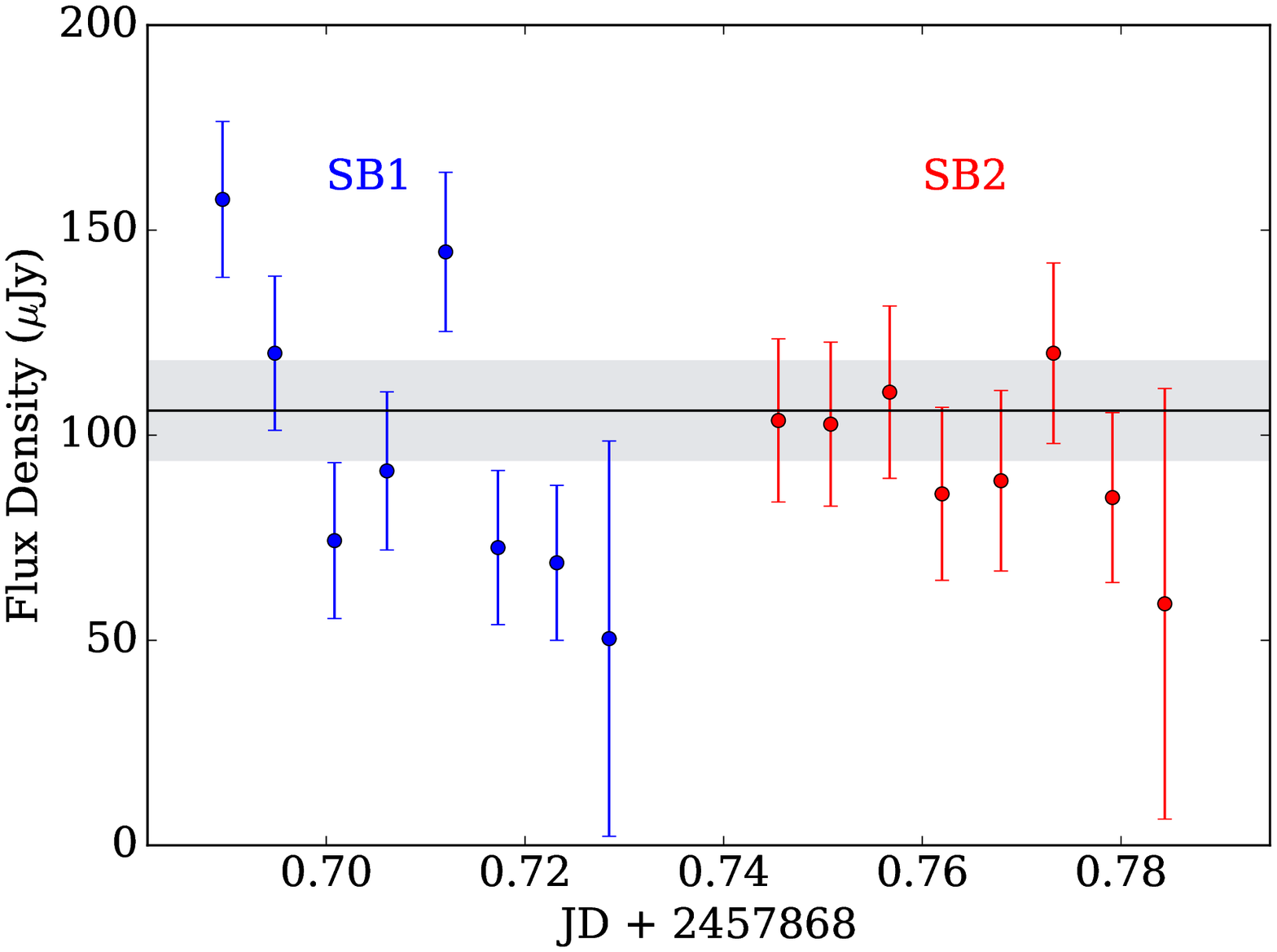}
  \end{center}
 \end{minipage}
\caption{\small Detection of millimeter emission from Proxima Centauri with the ALMA 12-m array. \emph{(left)} Natural weight images for all observations combined (left panel) along with each observation imaged separately (middle and right panels for SB~1 and SB~2, respectively).  In all images, contour levels are in steps of $[-2,2,4,6,...]\times$ the rms noise of 9.5, 13, and 14~$\mu$Jy~beam$^{-1}$ (left to right).  The stellar position is marked by the blue star symbol, and the $0\farcs99\times0\farcs87$ synthesized beam is indicated by the hatched ellipse in the lower left corner of the first panel.
\emph{(right)} Light curve of the detected point source over the course of both observations (SB~1 in blue and SB~2 in red).  The flux density was determined for each 6.58~minute scan using \texttt{uvmodelfit} in \texttt{CASA} to fit a point source model to the millimeter visibilities.  The solid black line and shaded gray region indicate the flux density of $106\pm12$~$\mu$Jy obtained by \cite{ang17} for comparison.  Millimeter emission is clearly detected from Proxima Centauri, with evidence for minor variability.  
}
\label{fig:12m}
\end{figure}

Figure~\ref{fig:12m} (leftmost panel) shows the natural weight continuum image produced by combining both ALMA 12-m array observations.  With natural weighting, the rms noise in the image is $9.5$~$\mu$Jy~beam$^{-1}$, and the beam size is $0\farcs99\times0\farcs87$ ($1.3\times1.1$~AU, position angle $=56\degr$).  An unresolved source is detected at the center of the image with a significance of $10\sigma$.  Fitting a point source model to the visibilities using the \texttt{CASA} task \texttt{uvmodelfit} yields a total flux density of $101\pm9$~$\mu$Jy, consistent with the result of $106\pm12$~$\mu$Jy reported by \cite{ang17}.  Any difference in flux density likely results from the fact that we fit to the millimeter visibilities, while \cite{ang17} perform all of their fits in the image domain.  The two middle panels of Figure~\ref{fig:12m} show the images that result from imaging each observation individually.  The rms noise is 13~$\mu$Jy~beam$^{-1}$ and 14~$\mu$Jy~beam$^{-1}$ for SB~1 and SB~2, respectively.  Both images show a central point source with measured flux density of $110\pm19$~$\mu$Jy and $94.1\pm8.0$~$\mu$Jy, respectively, consistent with each other within the uncertainties.  For comparison, the expected photospheric flux at 1.3~mm is $74\pm4$~$\mu$Jy, obtained by extrapolating the emission model derived by \cite{Ribas:2017} through a fit to the stellar Spectral Energy Distribution (SED).

Both ALMA 12-m SBs were executed on 2017 April 25 between 04:32:31.2--06:50:33.4 UTC.  A total of 16 (8 in each SB) on-source integrations or `scans' were performed during this period, each with a duration of 6.58~minutes.  We fit point source models to the millimeter visibilities in each scan independently to constrain any variability in the point source flux density.  The resulting light curve is shown in Figure~\ref{fig:12m} (right panel).  We note that the final scan in each SB was only 1.01~minutes in duration, resulting in substantially larger uncertainty.  There is some variation in flux density between successive scans, ranging between $157\pm19$~$\mu$Jy at maximum to $50.4\pm48.2$~$\mu$Jy at minimum.  The standard deviation for these data is $\sigma=29.1$~$\mu$Jy, and none of the individual scans have a flux density that differs from the mean by more than $3\sigma$.  We calculate the normalized `excess variance,' $\sigma_\text{rms}^2=\frac{1}{N\mu^2}\sum_{i=1}^{N}[(X_i-\mu)^2-\sigma_i^2]$ following \cite{Nadra:1997}, where $N$ is the number of data points, $\mu$ is the unweighted mean, and $X_i$ are the measured flux densities for each scan with errors $\sigma_i$. For these data, $\sigma_\text{rms}^2=0.017\pm0.030$. 

The observations include two linear polarizations (XX and YY) and span $\sim18$~GHz in total frequency between $224-242$~GHz, which allows us to determine a lower limit on the fractional linear polarization and a spectral index for the detected point source.  Combining both observations together, but imaging the XX and YY polarizations separately, yields total flux densities of $F_\text{XX} = 105\pm7$~$\mu$Jy and $F_\text{YY} = 96.2\pm10.0$~$\mu$Jy.  Given this, we can calculate the magnitude of the Stokes $Q=\langle E_\text{X}^2\rangle - \langle E_\text{Y}^2\rangle$ parameter as a fraction of the Stokes $I=\langle E_\text{X}^2\rangle+\langle E_\text{Y}^2\rangle$ to be $|Q/I|=0.09\pm0.12$, consistent with zero.  Since there are no polarization calibration and cross products available for these observations, $|Q/I|$ provides only a lower limit on the fractional linear polarization.  To calculate the spectral index, we determine the flux density of the point source detected in the combined image for the $225+227$~GHz and $239+241$~GHz spectral windows separately, yielding values of $100\pm13$~$\mu$Jy and $122\pm14$~$\mu$Jy, respectively.  The resulting spectral index is $\alpha = 2.58 \pm2.05$, where $F_\nu \propto \nu^\alpha$.  Within the significant errors, this result is consistent with blackbody emission in the Rayleigh-Jeans regime as expected for a quiescent stellar photosphere.

\subsection{Atacama Compact Array (ACA) Observations}
\label{sec:aca}

\begin{figure}[htbp]
\begin{minipage}[h]{0.55\textwidth}
  \begin{center}
  \vspace{0.5cm}
       \includegraphics[scale=0.55]{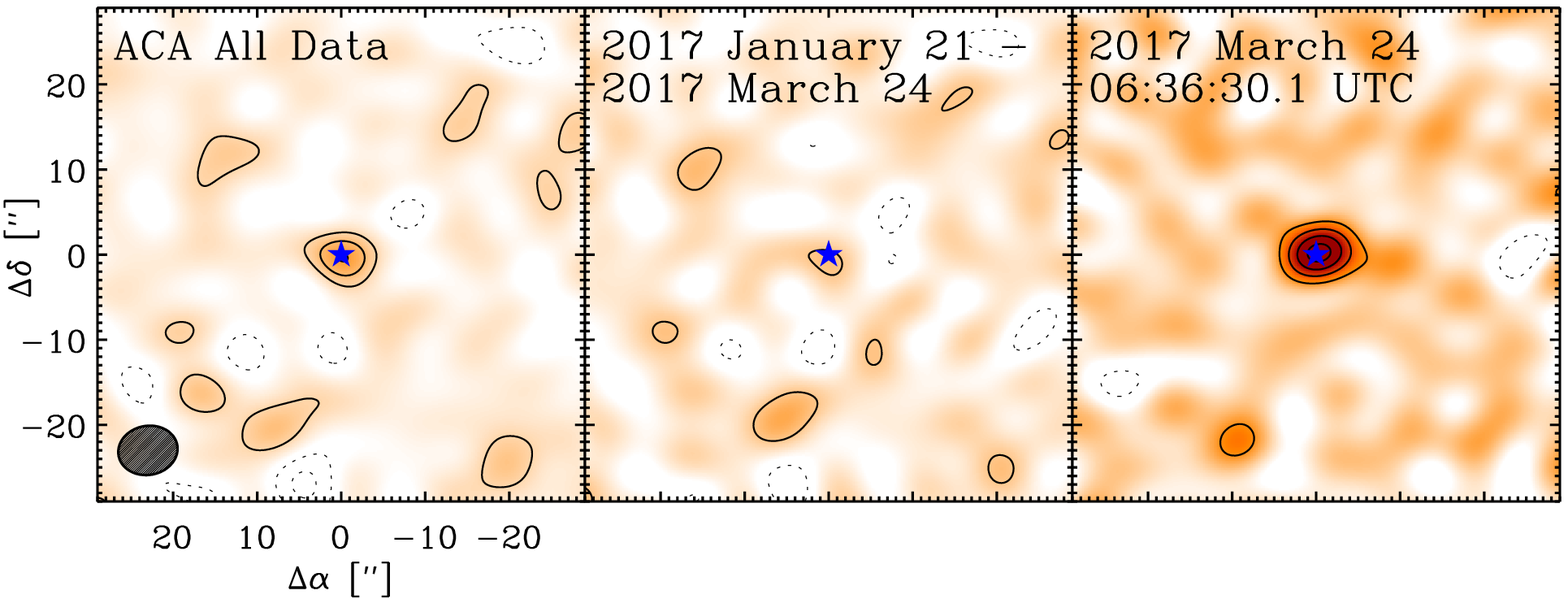}
  \end{center}
 \end{minipage}
\begin{minipage}[h]{0.5\textwidth}
  \begin{center}
       \includegraphics[scale=0.35]{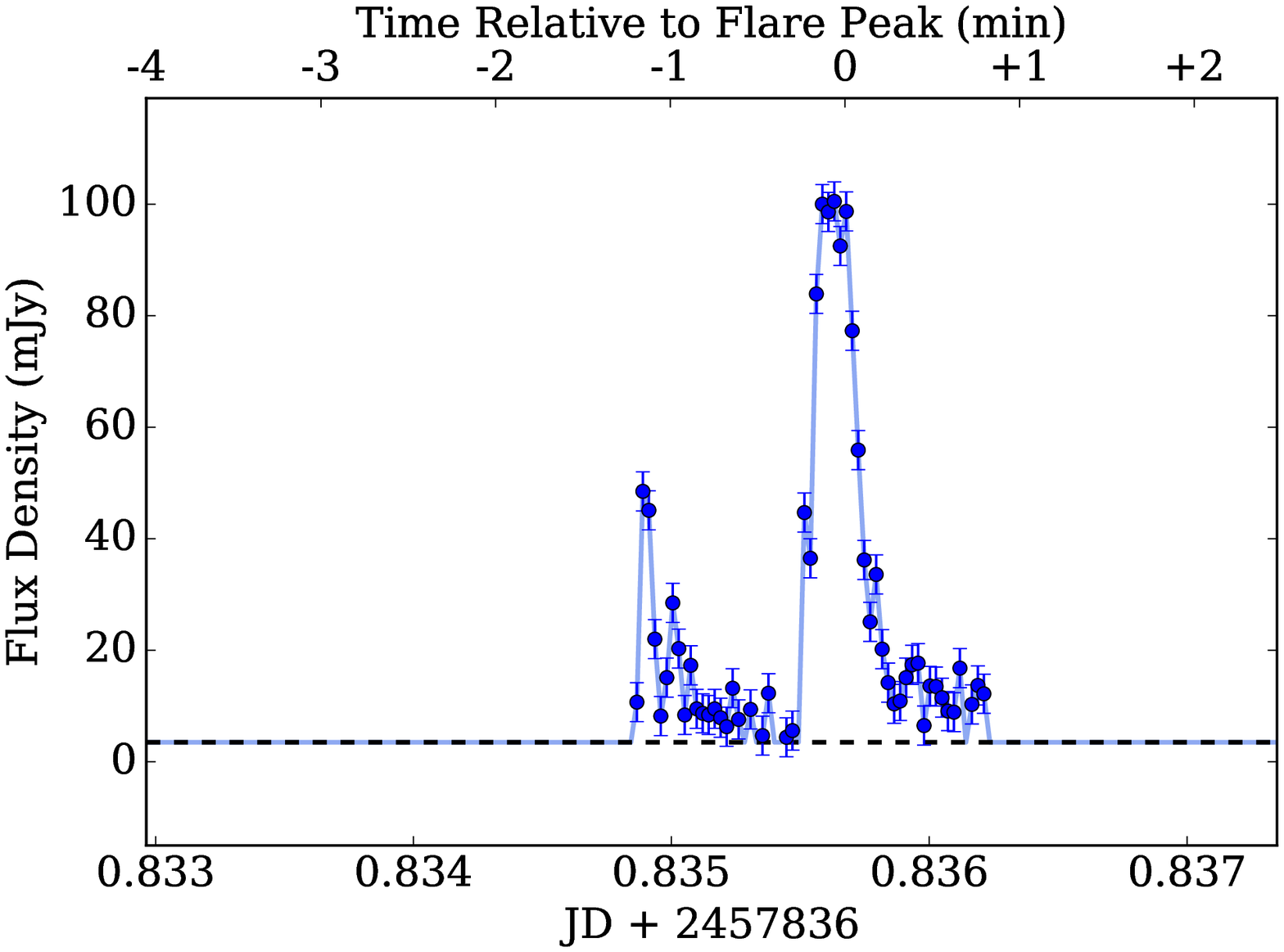}
  \end{center}
 \end{minipage}
\caption{\small Millimeter emission from Proxima Cen is detected only in a small subset of data taken with the ACA. \emph{(left)} The left panel shows the natural weight image that results from combining all 13 observations together.  The middle panel shows the first 12 observations combined, while the right panel shows the final observation imaged separately.  From these images, it is clear that a point source is only detected at the stellar location in the final observation.  In all three images, contour levels are in steps of $[-2,2,4,6,...]\times$ the rms noise of 47, 68, and 150~$\mu$Jy~beam$^{-1}$, respectively.  The stellar position is marked by the blue star symbol, and the $7\farcs27\times5\farcs51$ synthesized beam is indicated by the hatched ellipse in the lower left corner of the first panel.
\emph{(right)}  A light curve of the transient point source detected in the second to last scan of the observation on 2017 March 24 from 06:36:30.1--08:10:27.2 UTC.  Point source models were fit to the visibilities in 2~second intervals using \texttt{uvmodelfit} in \texttt{CASA}.  We only plot flux densities (blue points) for time intervals in which a central point source was detected.  For all other time intervals the flux density is less than the $3\sigma$ detection threshold of $3.5$~mJy indicated by the dashed black line.  A series of several small oscillations in flux density followed by a much larger flare-like event are seen at JD 2457836.8349--2457836.8355.
}
\label{fig:aca}
\end{figure}

Figure~\ref{fig:aca} (leftmost panel) shows the natural weight continuum image resulting from combining all 13 observations taken with the ACA between 2017 January 21 and 2017 March 24.  The rms noise is $47$~$\mu$Jy~beam$^{-1}$, and the beam size is $7\farcs3\times5\farcs5$ ($9.5\times7.2$~AU, position angle $=-79\degr$).  An unresolved point source with flux density $334\pm48$~$\mu$Jy ($7\sigma$) is detected, in agreement with the value of $340\pm60$~$\mu$Jy from \cite{ang17}, with the small difference again likely stemming from the choice in fitting to the visibilities rather than the image.  As with the 12-m data, we examined all 13 observations individually to explore the possibility of variability in the point source flux.  No sources above $3\sigma$ were detected in any of the first 12 observations (even when binned in 30 second intervals that should reveal short flares), and we show the natural weight image produced by averaging these datasets in Figure~\ref{fig:aca} (left middle panel).  The image is essentially noise-like; however, the $\sim2\sigma$ central peak could indicate the occurrence of small flares below the detection threshold of these observations.  Given the rms noise in this image, $68$~$\mu$Jy~beam$^{-1}$, a point source with the measured ALMA 12-m flux density of $\sim100$~$\mu$Jy would not have been detected. Figure~\ref{fig:aca} (right middle panel) shows an image of only the final observations executed on 2017 March 24 from 06:36:30.1--08:10:27.2 UTC with an rms noise level of $150$~$\mu$Jy~beam$^{-1}$.  In this final image, a bright central source is clearly seen at the stellar position with a flux density of $1.17\pm0.10$~mJy ($9\sigma$).

It is clear from this analysis that the point source seen in the combined ACA image is dominated by a transient event that occurred during the final observation on 2017 March 24.  During this complete observation, a total of 7 on-source scans of 6.58~minutes each were executed.  The central point source is only visible in the second to last scan.  In order to better constrain the time variable nature of this observed source, we fit models to the visibilities in this scan in 2~second intervals using \texttt{uvmodelfit} in \texttt{CASA}.  These fits are point-like with respect to the size of the synthesized beam, $7\farcs3\times5\farcs5$ ($9.5\times7.2$~AU).  The resulting light curve is shown in Figure~\ref{fig:aca} (right panel).  Several small peaks in flux density occur starting at JD 2457836.8349 followed by a steep rise in brightness at JD 2457836.8355 and an exponential fall-off.  This steep rise and more gradual fall-off is characteristic of flares from Proxima Centauri observed at other wavelengths \cite[e.g.,][]{Kowalski:2016}.  At peak brightness, the flux density reaches $100\pm4$~mJy or a luminosity of $2.04\pm0.15\times10^{14}$~erg~s$^{-1}$~Hz$^{-1}$ at 1.3~pc, a factor of nearly $1000\times$ brighter than during quiescence as measured in the ALMA 12-m observations. The complete transient event lasts for approximately 1~minute in total (FWHM of 30~seconds) before the source fades away below the noise level.  We do not detect any variability in the calibrator flux densities when imaged on similar timescales.

\begin{figure}[htbp]
  \begin{center}
  \vspace{0.5cm}
       \includegraphics[scale=0.52]{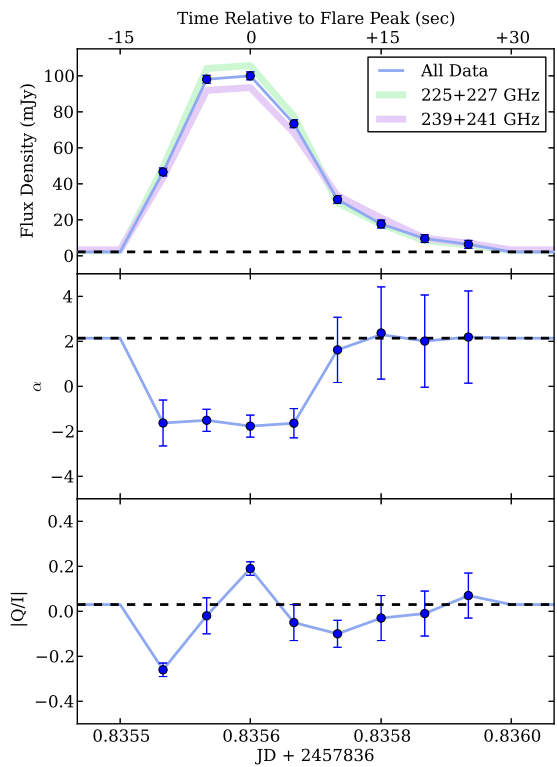}
  \end{center}
\caption{\small In addition to a rapid and large change in flux density, the flare emission has different spectral index and polarization compared to the quiescent emission.  The flux density \emph{(top)}, spectral index with frequency, $\alpha$, where $F_\nu \propto \nu^\alpha$ \emph{(middle)}, and lower limit on the fractional linear polarization ($|Q/I|$) \emph{(bottom)} during the observed stellar flare.  Point source models were fit to the visibilities in 5~second intervals using \texttt{uvmodelfit} in \texttt{CASA} to improve signal-to-noise.  The dashed line indicates the quiescent value of each parameter.  In the top panel, we plot the flux density of the $225+227$~GHz (green) and $239+241$~GHz (purple) spectral windows separately, along with the flux density for all spectral windows combined together (blue).
}
\label{fig:flare}
\end{figure}

As with the 12-m data, we can examine whether there is any change in the spectral index and polarization during the detected flare.  Figure~\ref{fig:flare} shows the flux density, spectral index as a function of frequency ($\alpha$), and lower limit on the fractional linear polarization ($|Q/I|$) during the course of the flare binned in 5~second intervals to improve signal-to-noise.  As above, we calculate the spectral index between the $225+227$~GHz and $239+241$~GHz spectral windows (plotted separately in Figure~\ref{fig:flare} as the green and purple thick lines, respectively). During the flare, the spectral index becomes negative, $F_\nu \propto \nu^\alpha$ with a minimum value of $\alpha = -1.77\pm0.45$ and an average value of $\alpha = -1.64\pm0.21$.  During the initial steep rise in flux density, the lower limit on the fractional linear polarization given by $|Q/I|=-0.26\pm0.02$.  At the peak, the sign flips to positive values, $|Q/I| = 0.19\pm0.02$, before returning to zero as the flux density declines. Both $\alpha$ and $|Q/I|$ differ significantly from the `quiescent' values determined from the 12-m ALMA data by $>5\sigma$, indicating a change in the source of the emission during the flaring event. Since we only have a lower limit, the origin of the changing $|Q/I|$ value is unclear from these observations.  \cite{Slee:2003} detected significant right circular polarization during a flare from Proxima Centauri at 1.38~GHz, which they attribute to either plasma emission or electron--cyclotron maser emission.  The signal-to-noise of the `pre-flare' at JD 2457836.8349 is not high enough to allow us to do the same analysis for this smaller event.

\section{Discussion}
\label{sec:discussion}

\subsection{Nature of the Detected Stellar Flare}
\label{sec:star}

Proxima Centauri is known to undergo frequent X-ray flares at the rate of one large event ($10\times$ quiescent level) every few days \citep{Haisch:1983,Gudel:2002,Gudel:2004,Fuhrmeister:2011,Kowalski:2016}. \cite{Gudel:2004} observed the star for 65~ks with XMM-Newton and caught a large X-ray flare with a peak luminosity of $3.9\times10^{28}$~erg~s$^{-1}$ and total X-ray energy of $1.5\times10^{32}$~erg that lasted for $\sim10$~minutes.  Low-level X-ray variability from small flares is observed continually.  Visual wavelength flares occur at a similar rate of several per day \citep{Davenport:2016}.  By extrapolating the flare energy frequency distribution, \cite{Davenport:2016} estimate that Proxima Centauri might display $\sim8$ visual flares each year with energy $>10^{33}$~erg.  

Stellar flares are typically associated with magnetic reconnection events. Late type stars such as Proxima Centauri are fully convective and have high surface magnetic fields \citep{Reiners:2008}.  X-ray observations suggest that Proxima Centauri's coronal flares are similar in temperature and density to those on the Sun, while also exhibiting the Neupert effect \citep{Fuhrmeister:2011}.  In the standard model of solar and stellar flares, a release of magnetic energy accelerates electrons that then impinge on hot ionized gas in the surrounding plasma. The flare produces emission across the entire electromagnetic spectrum: thermal X-rays from gas at $\sim$10$^7$~K, thermal bremsstrahllung (free-free) radiation at optical to radio wavelengths, and gyrosynchrotron emission in the radio. The observed slow decay depends on the cooling mechanism, and the timescale is typically set by the optical depth of the flare region. In principle, the shape and energy of the flare can be used to measure the emission mechanism, electron density, and size/optical depth of the emitting region. 

Flares from M dwarf stars have not previously been well-studied at (sub)millimeter wavelengths.  However, RS CVn binaries have been known to exhibit large millimeter flares \citep{Brown:2006,Beasley:1998,Massi:2006}.  Millimeter emission has also been detected in solar flares.  These events are associated with extreme X-ray emission (class X6 and above), have durations of minutes, and have positive spectral indices with frequency that range from 0.3--5 \citep{Krucker:2013}.  Even for the Sun, the origin of this emission is unclear since both free-free and gyrosynchrotron emission fail to reproduce the correct slope and X-ray fluxes given reasonable assumptions for the size and optical depth of the emission region.  Two possible emission mechanisms are synchrotron emission from ultra-relativistic electrons or an extreme-IR thermal extension of the heated chromosphere at 10,000~K.  The maximum (sub)millimeter flux density of a solar flare observed by \cite{Krucker:2013} is $\sim7\times10^8$~Jy corresponding to a luminosity of $\sim2\times10^{13}$~erg~s$^{-1}$~Hz$^{-1}$.  The peak luminosity of the observed Proxima Centauri flare, $2.04\pm0.15\times10^{14}$~erg~s$^{-1}$~Hz$^{-1}$, is a factor of $10\times$ larger.  Furthermore, the observed spectral index, $\alpha = -1.77\pm0.45$, is quite steep (modulo the significant uncertainty) and contradictory to the observations of solar flares that instead suggest a rising (positive) spectral index ($\alpha>2$).  This spectral index may indicate that we are seeing the optically thin part of the gyrosynchrotron spectrum, where the power-law index of non-thermal electrons, $\delta$, can be inferred from $\alpha$:  $\alpha = 1.22 - 0.9\delta$ \citep{Dulk:1985,Osten:2016}.  From the observed spectral index, we can determine $\delta=3.32\pm0.83$, within the range of $2.2\leq\delta\leq3.9$ expected for relatively hard radio spectra.  Given the anomalous nature of this flaring event and the fact that observations are only available in a narrow wavelength range, it is difficult to definitively constrain the nature of the emission mechanism.  Future observations at longer radio ($\sim1$~cm) wavelengths, near the expected peak of gyrosynchrotron emission, with the frequency resolution possible with the Australia Telescope Compact Array (ATCA) or ACA along with coincident X-ray measurements are needed to better understand this new flaring regime.

\subsection{New Constraints on Dust Emission}
\label{sec:dust}

\cite{ang17} propose a system of three dust belts to explain the ALMA 12-m and ACA observations of Proxima Centauri: (1) warm dust at $\sim0.4$~AU, (2) a cold belt from $1-4$~AU, and (3) an outer belt at $\sim30$~AU.  They note that the warm dust (1) and outer belt (3) are marginal detections, while only the cold belt (2) is detected with confidence. The warm dust is inferred from the $\sim30$~$\mu$Jy difference between the stellar photosphere and the total flux density of the 12-m image.  The cold belt is inferred from the $\sim200$~$\mu$Jy apparent flux difference between the expected photospheric contribution and the total flux density measured in the ACA image.  The outer belt position and geometry are determined by averaging the observed millimeter emission in elliptical annuli centered on the star.  Given that our reanalysis of the data reveals that Proxima Centauri experienced a significant flaring event during the ACA observations, the hypothesis of dust emission from at least belts (1) and (2) must be re-examined.

It is clear from our analysis that the entirety of the 1.3~mm emission detected by the ACA results from the short duration stellar flare.  No emission is detected above a $3\sigma$ level during quiescence (see Figure~\ref{fig:aca}).  This flare reaches a peak flux density of $100\pm4$~mJy, but lasts for only $\sim1$ minute.  When averaged together with the remaining $\sim19$~hours of non-detections, this transient event appeared as the $\sim340$~$\mu$Jy source detected by \cite{ang17}.  Without the detection of a quiescent excess, there is no need to posit the cold belt at $1-4$~AU. 

Given the high level of stellar activity displayed by Proxima Centauri, the $\sim30$~$\mu$Jy excess detected by the 12-m observations could also be plausibly explained by stellar emission.  The quiescent X-ray luminosity of the star varies by at least a factor of a few, and small X-ray and optical flares are observed frequently \citep{Fuhrmeister:2011,Gudel:2004}. Thus, it is reasonable to assume that Proxima Centauri may also undergo many small radio flares. If we assume that small flares have a flat spectrum at millimeter wavelengths, which likely overestimates the luminosity of the emission if it has the $\nu^2$ spectral index of thermal bremsstrahlung, then the 1.3~mm excess of $\sim30$~$\mu$Jy measured by the 12-m array yields a quiescent millimeter luminosity of 10$^{21}$~erg~s$^{-1}$ or less than $10^{-9}\times$ the bolometric luminosity of the star.  The $2\sigma$ peak seen in the first 12 ACA observations (Figure~\ref{fig:aca}) at the expected stellar position could also indicate some weak flare emission during these observations.  If the observed 12-m array excess is indeed due to low-level flaring activity, then there is no evidence for any warm dust emission at $\sim0.4$~AU.  Coronal emission at millimeter wavelengths has been observed for other M dwarf stars including AU Mic, which hosts a resolved cold debris disk at $\sim40$~AU and exhibits continual quiescent 1.3~mm emission in excess of expected photospheric levels and variable 9~mm emission \citep{mac16a}. Models of time-averaged turbulent coronal heating (which exhibits flare-like bursts in simulations) result in thermal bremsstrahlung sufficient to explain both the 9~mm variability and 1.3~mm emission, as well as the observed X-ray emission \citep{Cranmer:2013,mac16a}. The quiescent millimeter emission from AU Mic is considerably larger, however, with a 1.3~mm flux density 6$\times$ the photospheric level and 10$\times$ the excess observed from Proxima Centauri \citep{mac13}.  To support the claim of an inner warm disk, \cite{ang17} also note a slight elongation of the 12-m central source along a position angle of $130\degr$.  Our fits to the millimeter visibilities using \texttt{uvmodelfit} in \texttt{CASA} are consistent with a point source, and do not suggest that this source is spatially resolved. Any elongation of the central source is at a $2\sigma$ level in our natural weight images and varies between the two observations (see Figure~\ref{fig:12m}).    

The outer disk at $\sim30$~AU is posited by \cite{ang17} from several additional $\gtrsim3\sigma$ peaks in the ACA image that contribute to an azimuthally summed flux density of $\sim1.7$~mJy.  We find that the ACA visibilities do not adequately cover the short spatial frequencies needed to measure the total disk flux density and make it impossible to robustly constrain the disk geometry.  Additional observations are needed to test the veracity of this detection.  The possibility of contamination must also be considered.  Each of the peaks that contribute to the putative disk have a flux density of  $\sim150-200$~$\mu$Jy.  Deep ALMA surveys at millimeter wavelengths have placed strong constraints on the expected number of faint background sources in an image \citep{Carniani:2015} that agree with results from cosmological simulations \citep{Shimizu:2012,Cai:2013}.  Given the differential source counts at 1.3~mm, the number of sources with flux density $>150$~$\mu$Jy that are expected within the ACA primary beam (FWHM$\sim39\arcsec$) is $13^{+10}_{-8}$, a not insignificant number.  In addition, Proxima Centauri sits close to the Galactic plane (latitude $-2\degr$) in a region of high background cirrus, which made it impossible for Spitzer to provide an upper limit on the source flux density at 60~$\mu$m \citep{Gautier:2007}. Between possible extragalactic background and Galactic sources, it will be difficult to detect or image the proposed extended disk.

\section{Conclusions}
\label{sec:conclusions}

We present new analyses of ALMA 12-m and ACA observations at 233~GHz (1.3~mm) of the Proxima Centauri system.  The ALMA 12-m observations reveal a central emission peak consistent with a point source with flux density $101\pm9$~$\mu$Jy.  During the ACA observations, the star underwent a significant flaring event, reaching a peak flux density of $100\pm4$~mJy, nearly a factor of $1000\times$ brighter than during the 12-m observations.  At the flare peak, the 1.3~mm continuum emission is characterized by a steeply falling spectral index with frequency, $F_\nu\propto \nu^\alpha$ with $\alpha=-1.77\pm0.45$, and a lower limit on the fractional linear polarization, $|Q/I|=0.19\pm0.02$.  This detection of an M dwarf flare at millimeter wavelengths, opens a new observational window on the mechanisms responsible for stellar flares.  Additional observations at radio through X-ray wavelengths of M dwarf flares from Proxima Centauri and other systems are needed to better understand the properties of such flares and their implications for the habitability of exoplanets in these systems.

The quiescent emission detected by the sensitive 12-m array observations lies below the detection threshold of the ACA observations, and the only ACA detection of Proxima Centauri is during a series of small flares followed by a stronger flare of $\sim1$ minute duration.  Due to the clear transient nature of this event, we conclude that there is no need to invoke the presence of an inner dust belt at $1-4$~AU.  It is also likely that the slight excess above the expected photosphere observed in the 12-m observations is due to coronal heating from continual smaller flares, as is seen for AU Mic, another active M dwarf that hosts a well-resolved debris disk.  If that is the case, then the need to include warm dust emission at $\sim0.4$~AU is removed.  Although the detection of a flare does not immediately impact the claim of an outer belt at $\sim30$~AU, the significant number of background sources expected in the image and known high level of background cirrus suggest that caution should be used in over-interpreting this marginal result.

\vspace{1cm}
The authors thank the anonymous referee for helpful suggestions that improved this manuscript.  M.A.M. acknowledges support from a National Science Foundation Astronomy and Astrophysics Postdoctoral Fellowship under Award No. 1701406. This paper makes use of the following ALMA data: ADS/JAO.ALMA \#2016.A.00013.S. ALMA is a partnership of ESO (representing its member states), NSF (USA) and NINS (Japan), together with NRC (Canada) and NSC and ASIAA (Taiwan) and KASI (Republic of Korea), in cooperation with the Republic of Chile. The Joint ALMA Observatory is operated by ESO, AUI/NRAO and NAOJ. The National Radio Astronomy Observatory is a facility of the National Science Foundation operated under cooperative agreement by Associated Universities, Inc.


\begin{thebibliography}{dummy}

\bibitem[Anglada et al.(2017)]{ang17} Anglada, G., Amado, P. J., Ortiz, J. L., et al.\ 2017, ApJL, 850, L6

\bibitem[Anglada-Escud{\'e} et al.(2016)]{ang16} Anglada-Escud{\'e}, G., Amado, P. J., Barnes, J., et al.\ 2016, Nature, 536, 437

\bibitem[Beasley \& Bastian(1998)]{Beasley:1998} Beasley, A. J. \& Bastian, T. S.\ 1998, in Astronomical Society of the Pacific Conference Series, Vol. 144, IAU Colloq. 164: Radio Emission from Galactic and Extragalactic Compact Sources, ed. J. A. Zensus, G. B. Taylor, \& J. M. Wrobel, 321

\bibitem[Brown \& Brown(2006)]{Brown:2006} Brown, J. M. \& Brown, A.\ 2006, ApJL, 638, L37

\bibitem[Cai et al.(2013)]{Cai:2013} Cai, Z.-Y., Lapi, A., Xia, J.-Q., et al.\ 2013, ApJ, 768, 21

\bibitem[Carniani et al.(2015)]{Carniani:2015} Carniani, S., Maiolino, R., De Zotti, G., et al.\ 2015, A\&A, 584, A78

\bibitem[Cranmer et al.(2013)]{Cranmer:2013} Cranmer, S. R., Wilner, D. J., \& MacGregor, M. A.\ 2013, ApJ, 772, 149

\bibitem[Davenport et al.(2016)]{Davenport:2016} Davenport, J. R. A., Kipping, D. M., Sasselov, D., Matthews, J. M., \& Cameron, C.\ 2016, ApJL, 829, L31

\bibitem[Dong et al.(2017)]{Dong:2017} Dong, C., Lingam, M., Ma, Y., \& Cohen, O.\ 2017, ApJL, 837, L26

\bibitem[Dressing \& Charbonneau(2015)]{Dressing:2015} Dressing, C. D. \& Charbonneau, D.\ 2015, ApJ, 807, 45

\bibitem[Dulk(1985)]{Dulk:1985} Dulk, G. A. 1985, ARA\&A, 23, 169

\bibitem[France et al.(2016)]{France:2016} France, K., Parke Loyd, R. O., Youngblood, A., et al.\ 2016, ApJ, 820, 89

\bibitem[Fuhrmeister et al.(2011)]{Fuhrmeister:2011} Fuhrmeister, B., Lalitha, S., Poppenhaeger, K., et al.\ 2011, A\&A, 534, A133

\bibitem[Garraffo et al.(2016)]{Garraffo:2016} Garraffo, C., Drake, J. J., \& Cohen, O.\ 2016, ApJL, 833, L4

\bibitem[Gautier et al.(2007)]{Gautier:2007} Gautier, T. N., Rieke, G. H., Stansberry, J., et al.\ 2007, ApJ, 667, 527

\bibitem[G{\"u}del et al.(2004)]{Gudel:2004} G{\"u}del, M., Audard, M., Reale, F., Skinner, S. L., \& Linsky, J. L.\ 2004, A\&A, 416, 713

\bibitem[G{\"u}del et al.(2002)]{Gudel:2002} G{\"u}del, M., Audard, M., Skinner, S. L., \& Horvath, M. I.\ 2002, ApJL, 580, L73

\bibitem[Haisch et al.(1983)]{Haisch:1983} Haisch, B. M., Linsky, J. L., Bornmann, P. L., et al.\ 1983, ApJ, 267, 280

\bibitem[Hawley et al.(1997)]{Hawley:1997} Hawley, S. L., Gizis, J. E., \& Reid, N. I.\ 1997, AJ, 113, 1458

\bibitem[Henry et al.(2006)]{Henry:2006} Henry, T. J., Jao, W.-C., Subasavage, J. P., et al.\ 2006, AJ, 132, 2360

\bibitem[Kowalski et al.(2016)]{Kowalski:2016} Kowalski, A. F., Mathioudakis, M., Hawley, S. L., et al.\ 2016, ApJ, 820, 95

\bibitem[Krucker et al.(2013)]{Krucker:2013} Krucker, S., Gim{\'e}nez de Castro, C. G., Hudson, H. S., et al.\ 2013, A\&A Rv, 21, 58

\bibitem[Li et al.(2017)]{Li:2017} Li, Y., Stefansson, G., Robertson, P., et al.\ 2017, ArXiv e-prints, arXiv:1712.04483

\bibitem[Lingam \& Loeb(2017)]{Lingam:2017} Lingam, M. \& Loeb, A.\ 2017, ApJL, 846, L21

\bibitem[MacGregor et al.(2013)]{mac13} MacGregor, M.~A., Wilner, D.~J., Rosenfeld, K.~A., et al.\ 2013, ApJ, 762, L21

\bibitem[MacGregor et al.(2016)]{mac16a} MacGregor, M.~A., Wilner, D.~J., Chandler, C., et al.\ 2016, ApJ, 823, 79 

\bibitem[Massi et al.(2006)]{Massi:2006} Massi, M., Forbrich, J., Menten, K. M., et al.\ 2006, A\&A, 453, 959

\bibitem[Nadra et al.(1997)]{Nadra:1997} Nadra, K., George, I. M., Mushotzky, R. F., et al.\ 1997, ApJ, 476, 70

\bibitem[Osten et al.(2016)]{Osten:2016} Osten, R. A., Kowalski, A., Drake, S. A., et al.\ 2016, ApJ, 832, 174

\bibitem[Petry \& CASA Development Team(2012)]{Petry:2012} Petry, D. \& CASA Development Team.\ 2012, in Astronomical Society of the Pacific Conference Series, Vol. 461, Astronomical Data Analysis Software and Systems XXI, ed. P. Ballester, D. Egret, \& N. P. F. Lorente, 849

\bibitem[Reiners \& Basri(2008)]{Reiners:2008} Reiners, A. \& Basri, G.\ 2008, A\&A, 489, L45

\bibitem[Ribas et al.(2017)]{Ribas:2017} Ribas, I., Gregg, M. D., Boyajian, T. S., \& Bolmont, E.\ 2017, A\&A, 603, A58

\bibitem[Shkolnik et al.(2014)]{Shkolnik:2014} Shkolnik, E. L. \& Barman, T. S.\ 2014, AJ, 148, 64

\bibitem[Shimizu et al.(2012)]{Shimizu:2012} Shimizu, I., Yoshida, N., \& Okamoto, T.\ 2012, MNRAS, 427, 2866

\bibitem[Slee et al.(2003)]{Slee:2003} Slee, O. B., Willes, A. J., \& Robinson, R. D.\ 2003, PASA, 20, 257

\bibitem[Thackeray(1950)]{Thackeray:1950} Thackeray, A. D.\ 1950, Monthly Notes of the Astronomical Society of South Africa, 9, 9

\bibitem[van Leeuwen(2007)]{van07} van Leeuwen, F.\ 2007, A\&A, 474, 653

\bibitem[Zahnle \& Catling(2017)]{Zahnle:2017} Zahnle, K. J. \& Catling, D. C. 2017, ApJ, 843, 122

\end{thebibliography}
\end{document}